# Metal-Oxide Sensor Array for Selective Gas Detection in Mixtures

*Noureddine Tayebi[†]\*, Varvara Kollia[†]\* and Pradyumna S. Singh\**

\*Intel Labs, Intel Corporation, 2200 Mission College Boulevard, Santa Clara, CA 95054, USA

ntayebi@alumni.stanford.edu, varvara.kollia@gmail.com, pradyumna.s.singh@intel.com

([†] Work reported herein was done when these authors were at Intel.)

**Abstract**

We present a monolithic, microfabricated, metal-oxide semiconductor (MOS) sensor array in conjunction with a machine learning algorithm to determine unique fingerprints of individual gases within homogenous mixtures. The array comprises four different metal oxides and is engineered for independent temperature control and readout from each individual pixel in a multiplexed fashion. The sensor pixels are designed on a very thin membrane to minimize heat dissipation, thereby significantly lowering the overall power consumption (<30µW average power). The high dimensional data obtained by running the pixels at different temperatures, is used to train our machine learning algorithm with an average accuracy ~ 88% for high resolution detection and estimation of concentration of individual constituents in a homogenous mixture. While the response of MOS sensors to various gases has been demonstrated, very few studies have investigated the response of these sensors to homogeneous mixtures of gases comprising several gases. We demonstrate this principle for a binary homogeneous mixture of ozone and carbon monoxide, both of which are criteria pollutant gases. Our findings indicate that a multiplicity of MOS elements together with the ability to vary and measure at various temperatures are essential in predicting concentration of individual gases within mixtures, thereby overcoming a key limitation of MOS sensors – poor selectivity. The small form-factor and microfabrication approach of our sensor array also lends itself to CMOS integration paving the way for a platform for wearable and portable applications.

# INTRODUCTION

There is much current interest in the development of miniaturized, low-power sensors for the selective and sensitive detection of pollutant gases. The World Health Organization attributed approximately 3.7 million deaths due to ambient (outdoor) air pollution and another 4.3 million deaths due to indoor air pollution in 2012.[1] Poor air quality has also been implicated as a contributing factor to increased incidences of cancer, respiratory, cardiovascular and neurovascular diseases.[2-6] Current methods of monitoring air quality rely on only a few monitoring stations in a large geographical areas that span cities and towns. While they can be accurate, they often fail to provide direct information regarding air quality in highly localized microenvironments.

By contrast, small, low-power sensors can enable individuals to monitor their own personal exposure to various gases ranging from the so-called "criteria" gases found in ambient air and regulated by the EPA to volatile organic compounds (e.g., benzene, formaldehyde) that are found in indoor or industrial settings. Such affordable, customizable gas sensors also have the potential to directly impact health outcomes by also reporting the health implications of the air quality in the immediate vicinity of vulnerable populations, e.g., people with respiratory ailments like asthma, chronic obstructive pulmonary disorder, and heart disease.

Furthermore, these low-cost, low-power sensors could serve as nodes in dense wireless sensor networks,[7,8] which can augment existing remote sensing techniques and transport models to offer better resolved spatial air pollution estimates.[9] Such networks have recently shown to identify sources of specific pollutants from specific sources, thus imparting the ability for smart interventions in the control of air pollution.[10]

Several sensor technologies exist today for the selective and sensitive detection of pollutant gases and volatile organic compounds (VOCs). These technologies differ in terms of the materials they employ (conducting polymers, dielectric polymers, metal-oxide semiconductors etc.) the properties they measure for signal transduction (capacitance, resistance, electrochemical current, optical absorption, fluorescence etc.) and form factors.[11, 12] Optical sensors via modalities as infrared (IR), Raman, UV-Vis absorption and fluorescence are extremely selective. Together with the canonical gas chromatography-mass spectrometry (GC-MS), they are the standard instrumental method for gas detection and are the basis of most bench-top level gas analyzers.[12] However, owing to the bulky, cumbersome, expensive and power-hungry nature of optical components involved, they are very difficult to miniaturize. Electrochemical sensors are also very selective towards certain gases (e.g., carbon monoxide, CO) but because the electrochemical current scales with area of the electrode, it is difficult to miniaturize them to ultra-small (~ mm scale) form factors, especially if they are to be components in e.g., mobile devices.[13] By contrast resistive methods, which rely on the change in the bulk resistance of conductive polymers[14] or heated metal-oxide semiconductors (MOS)[15], have been shown to be very sensitive to pollutant gases by routinely being able to detect ppb (parts-per-billion) levels of gases. Importantly, they are also scalable to very ultra-small form factors.

However, existing MOS-based sensors which typically utilize a single MOS material (typically Tix Oxide, SnO2) as the sensing element suffer from several limitations. Most MOS materials show cross-sensitivity to multiple gases and VOCs. Thus, selective determination of one or a few gases in a homogeneous mixture is very challenging using these sensors. While there have been many attempts to achieve selectivity though the doping of metal oxide films with additives such as Pt, Pd, Au, Ag, Cu etc., such solutions only provide partial selectivity increases.[16] Moreover, since the MOS material requires heating to be adequately sensitive, these sensors are also very power-hungry, with average power consumption in 10s of mW (milli Watt) range. Finally, their response is very susceptible to other environmental parameters, most notably fluctuations in humidity.

To mitigate the above mentioned limitations, we propose herein a monolithic, chip-based array of MOS materials wherein, each sensor element of the array– referred to as a pixel – is made of a different metal oxide. Each pixel is individually addressable and integrated with its own dedicated heater, temperature sensor and sensing electrodes (Fig. 1a.). Thus the temperature at each pixel can be independently controlled and resistance changes at the pixel read out individually. The use of an array relaxes the requirement for each individual pixel to be maximally selective to any one gas. Instead, following the idea of electronic nose[11, 17], it is the cumulative response from the multiplicity of different pixels, at differing temperatures, that is of interest. This composite response across all the pixels (Fig. 1b) which is fed into and used to train a hybrid machine learning algorithm to determine a unique "fingerprint" for precise gas detection and concentration determination. Arrays of MOS sensors together with pattern recognition techniques have been used in the past for a variety of gas sensing applications, ranging from sub-ppm (parts per million) detection of VOCs[18], combustion gases[19] and explosives[20]. However, all these studies involved the use of discrete, commercially available MOS sensors or sensing materials. To our knowledge, our devices are the first instance where MOS arrays of multiple MOS materials have been microfabricated to create a monolithic, chip-scale solution.

The paper is organized as follows. We begin the ensuing sections by describing the fabrication, and operational features of our device. We discuss the response of the array to individual gases as well as discuss results on homogeneous mixtures of gases, in particular ozone ($O_3$) and carbon monoxide (CO). We then describe our hybrid pattern recognition scheme. Finally, we conclude by summarizing the results and discussing the prospects of integrating these devices with standard CMOS process technology to create fully integrated, miniaturized, low-power solutions for selective gas detection.

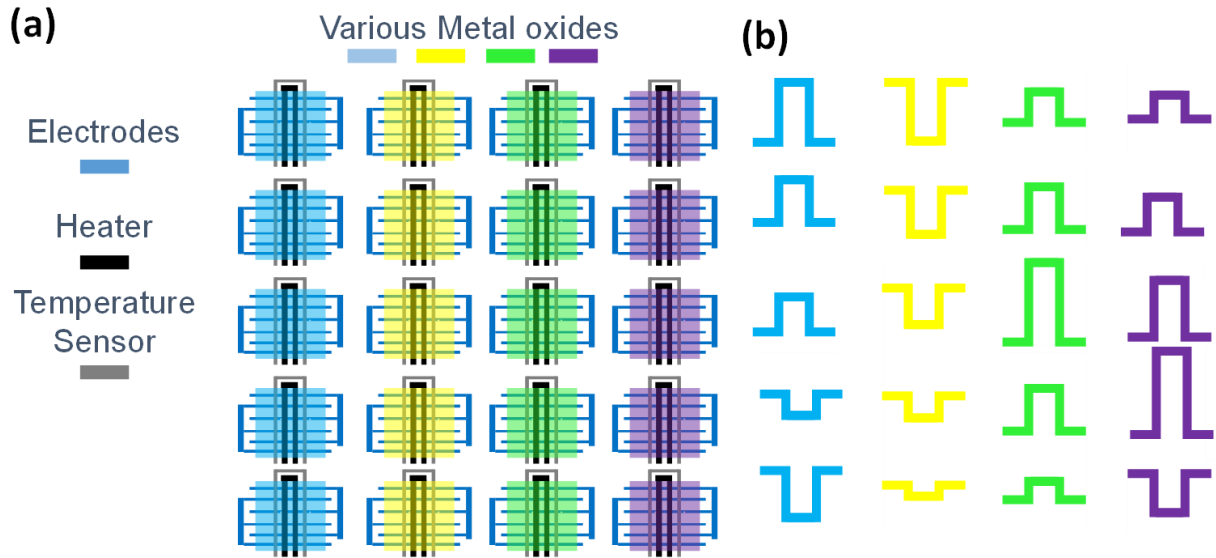

**Fig. 1.** Schematic of MOS sensor array principle. **(a)** Multiple array pixels of various metal oxides, simultaneously operating at various temperatures. **(b)** Schematic of response matrix from the array pixels used to train the machine learning algorithm used to achieve selective gas detection.

**MATERIALS AND METHODS**

We designed and fabricated a 3×2 sensor array using standard microfabrication techniques. The individual pixels are comprised of one out of a set of four MOS materials: $In_2O_3$, $SnO_2$, ZnO and $WO_3$. Two pixels in the array are thus redundant, but are used as test structures. Each sensor pixel consists of a free-standing 500 nm thick silicon nitride membrane (100×100 µm²) that serves as a substrate for the subsequent layers. 100 nm thick platinum layer of serpentine features is deposited which serves as heating and resistive (for temperature sensing) elements. This layer is passivated by a 200 nm silicon nitride film. This is followed by patterning 100 nm thick platinum interdigitated electrodes on top of which the various metal oxides are deposited with a 100 nm thickness. The silicon nitride film was deposited using low-pressure chemical vapor deposition at 250 mTorr engineered to result in very low stress membranes. This is critical to lower susceptibility to damage by repeated temperature cycling. Figure 2a shows the active area, whereas Figures 2b-c are zoom-in and cross-sectional views of a sensor pixel within the array. The pixels were optimized using finite element analysis to reduce the heat dissipation within the active area, which leads to a uniform temperature distribution that drops down to room temperature closer to the substrate (Figure 2d). Figure 2e shows the required power consumption to reach a certain temperature measured by the resistance change of the temperature sensors. For example, at 300 °C, the peak power is only 2 mW. This low power is due to the reduced heat dissipation favored by the thin silicon nitride membrane. Under pulse width modulation wherein sampling time is 2 seconds performed every 5 minutes during a 24-hour period, the average power consumption at 300 °C is only 13.5 µW.

We custom built a gas testing chamber to test the devices under varying gas concentrations. Carbon Monoxide (CO) gas was delivered to the device using CO cylinders (GASCO, Calibration gas, 50 ppm in Air). Ozone was generated using an Ozone generator (Analytical Instruments, PA) connected to a 20.9% O2/N2 cylinder (GASCO, Calibration gas). The ozone gas was delivered at a fixed flow rate of 500 SCCM. Oil-free Air was used for dilution purposes. The desired concentrations were set by controlling the flow rates of the individual gases using mass flow controllers (Brooks Instruments, GF series and 4800 series).The devices were addressed using a custom designed printed circuit board (PCB) connected to a Keithley 4200 Semiconductor analyzer.

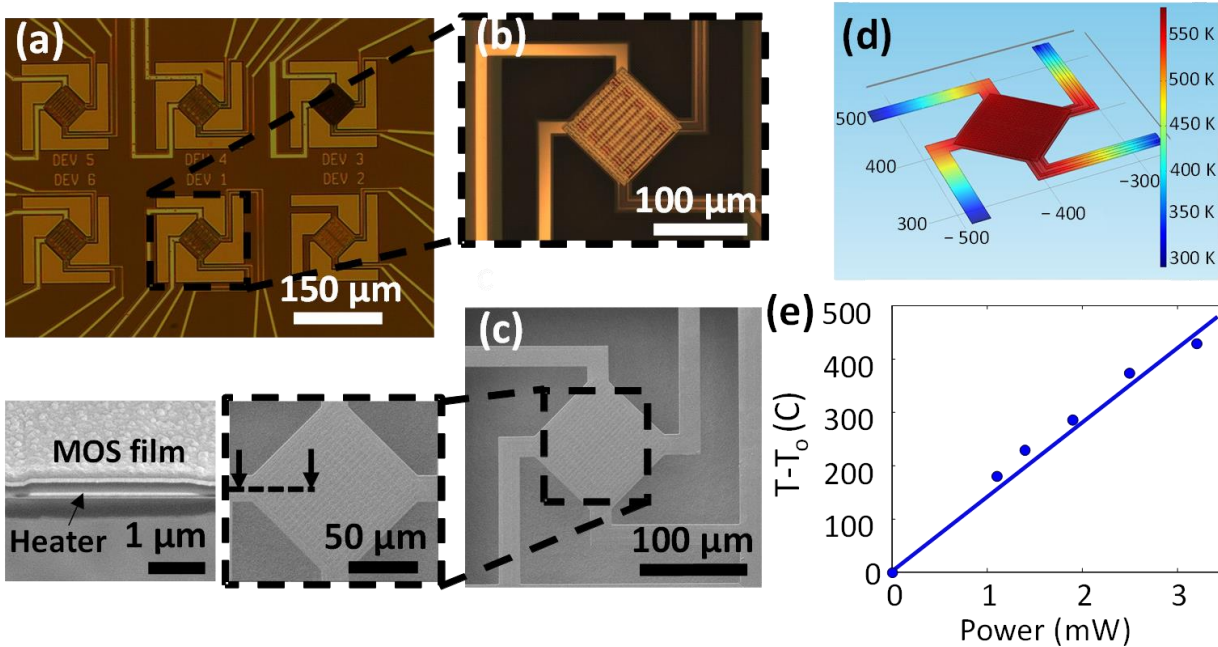

**Fig. 2.** MOS sensor array. **(a)** Optical image of the active area of the 3x2 sensor array wherein a different metal oxide is deposited on each pixel (In$_2$O$_3$ and SnO$_2$ are redundant pixels). **(b)** Zoom-in view of a single pixel showing the suspended membrane. The top layer corresponds to the metal oxide film under which interdigitated electrodes are used to measure the change of resistance in the presence of a gas. **(c)** Scanning electron microscopy images of a single pixel with a cross section that shows the various layers of the sensor that consist of the heater and temperature electrodes (sandwich film) and the metal oxide (top film) under which is the electrode. **(d)** Finite element simulation showing the uniform temperature distribution within the metal oxide film. **(e)** Variation of peak power with temperature. $T_o$ is the controlled ambient temperature of 22°C.

## RESULTS

The response of the MOS array was investigated for two gases, ozone (O$_3$) and carbon monoxide (CO), as well as for homogeneous mixtures of these two gases. The reason for selecting these two gases was their widely differing molecular properties. It is well known that O$_3$ is an oxidizing gas, while CO is a reducing gas. Differing molecular properties offers a good starting point for being able to resolve component gases in a mixture. Furthermore, our initial experiments are with a binary mixture for simplicity. The tests were performed in a custom-designed gas test set-up to introduce various test gases to the device in a controlled manner.

We first determined the optimum operating temperature for each metal oxide and for a given gas, by varying the temperature for each array pixel at various gas concentrations (0-800 ppb for $O_3$ and 0 to 50 ppm for CO in the presence of air). In the case of $O_3$, the highest response was obtained at 300 °C for $In_2O_3$, ZnO and $WO_3$ films and 200 °C for $SnO_2$. Figure 3a shows the change in resistance, determined as the ratio between the resistances in the presence and absence of $O_3$ (absence here refers to 100% air) with respect to $O_3$ concentration at the optimum temperature for each metal oxide. The change in resistance is clearly linear in the range of 30-800 ppb which spans the healthy, unhealthy, and hazardous ranges documented by the US environmental protection agency.[21] Note that $O_3$ is an oxidizing gas that induces an increase in resistance.

For the case of CO, the optimum temperature was obtained at 200 °C for $SnO_2$, ZnO and $WO_3$ films whereas $In_2O_3$ showed very low sensitivity to CO, which is in agreement with previous studies, wherein reducing gases tend to react with $In_2O_3$ at much higher temperature.[22] This can be taken advantage of in the fingerprinting scheme to isolate the presence (or absence) of CO in a given mixture. Linear variation in the response to CO concentrations is also observed as shown in Fig. 3b. In this case, the change in resistance is defined as the ratio between the absence and presence of CO in air, as CO is a reducing gas which induces a decrease in resistance.

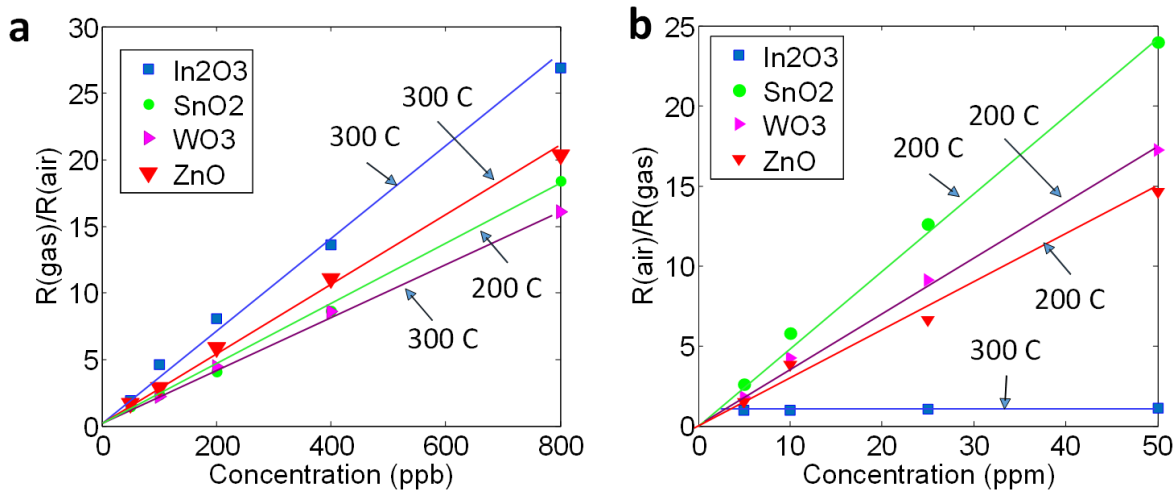

**Fig. 3.** Change of resistance as a function of concentration at optimum (highest sensitivity) temperatures for **(a)** $O_3$ and **(b)** CO; since exposure to CO results in a decrease in resistance, $R(air)/R(gas)$ is used. Note that there was no reaction to CO by the $In_2O_3$ pixel in the 200-400°C temperature range. Thus shown here is the 300°C temperature only.

Once the trends for the individual gases were obtained, binary homogeneous $O_3$-CO mixtures were introduced at various concentration combinations and measured at various temperatures for each metal oxide pixel. 10 concentrations of $O_3$ (0-800 ppb) and 9 concentrations of CO (0-50 ppm) were chosen resulting in 90 unique concentration combinations. Figure 4a is a 3D histogram which shows the resistance change (z-axis) at a given concentration combination (x-axis for $O_3$ and y-axis for CO) for the case of $SnO_2$ at 200° C. For clarity, only 5 concentrations for each gas are shown. The negative values correspond to a resistance reduction (case when CO signal dominates $O_3$ signal), whereas positive values correspond to a resistance increase (case when $O_3$ signal dominates CO signal). Similar data sets were obtained for all the metal oxide pixels and at various operating temperatures.

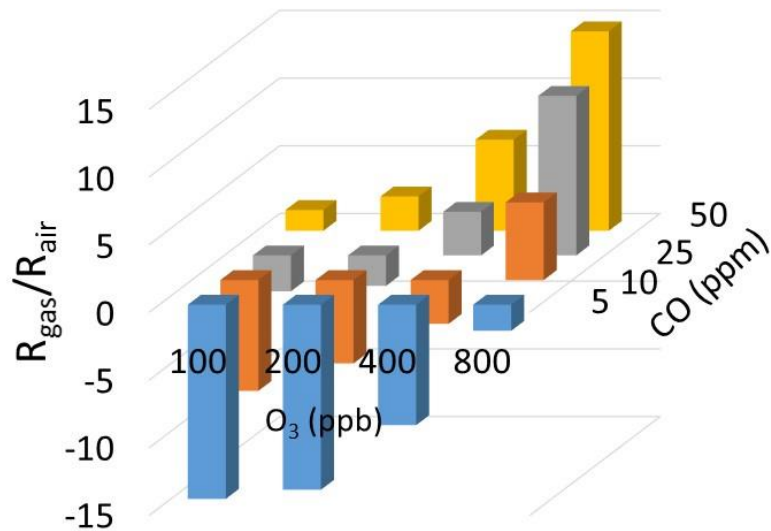

**Fig. 4.** 3D histogram of resistance change (z-axis) at a given concentration combination (x-axis for $O_3$ and y-axis for CO) for the case of $SnO_2$ at 200°C. Similar data sets (not shown) were also obtained for various metal oxides and operating temperatures.

## DISCUSSION

**Data Analysis**

As there is increasing interest in the field of gas array sensing, there is matching increased interest in algorithmic analysis, to enhance the gas detection performance. A good overview of the state-of-the- art pattern analysis techniques in this field is given in the review by R. G. Osuna.[23] Most of the research is focused on the detection of independent gases, however there has also been recently increasing research in the field of gas-mixture identification.[24, 25] However, these studies, to the best of the authors' knowledge, do not pertain to the use of temperature as a pivotal element in the definition of the concentration ranges of the gases in question, which is one of the main elements in our pattern analysis.

The problem of gas-mixture identification can be posed as either an unsupervised one[26], where patterns in the data are detected via cluster identification or self-organizing maps, or it can be solved in a supervised fashion[27], where the input resistances are used to train a detection model. In our case, we will work in a supervised setting.

To enhance the detection mechanism, different operating temperatures are frequently used. In fact, temperature modulation schemes have been reported to correspond to different response patterns.[28] Here, we focus our analysis on data collected at three different temperatures: 200° C, 300° C and 400° C. In particular, the experimental data were taken from varying both the $O_3$ and the CO concentrations simultaneously within the mixture, with corresponding variations in ppb and ppm ranges, respectively. The four variables of the input correspond to the normalized resistance values of the 4 metal-oxide materials used ($M_1$-$M_4$), namely $In_2O_3$, $SnO_2$, $WO_3$ and ZnO. The normalized resistance is defined as the ratio of each metal oxide resistance $R$ in the presence of the mixture over its reference, which is the baseline resistance in the presence of air only. This is referred to as $R_{ref}$. All the values are measured at a given temperature $T$. The input variables correspond to mean values over the detection period of 4 seconds. These input variables are subsequently processed by a machine learning pattern recognition algorithm. The final goal for the algorithm is to predict the concentration of each component gas at each of the unique mixture combinations, based on these inputs. Here $Gas_1$ and $Gas_2$ refer to $O_3$ and CO, respectively.

**Hybrid Multi-Stage Algorithm**

We employ a hybrid machine learning algorithm to detect each gas' concentration in the mixture. The underlying idea is that different temperature ranges can be used to detect different gas' concentrations. Multiple linear regression models[29, 30] are used to fit different concentrations ranges of different gases within a mixture, where each model corresponds to data from a distinct temperature. In the case that the regression model cannot fit the data, a more elaborate technique (artificial neural network[29]) is employed to augment the classifier performance. We will now give a brief overview of the two algorithms we combine in the multiple stages of our hybrid method. Figure 5a shows the schematic set-up of the pattern recognition problem.

The basic mechanism behind linear regression[29, 31] is as follows. If we denote our predictor variable with y and our input vector x, estimating y from x with linear regression is equivalent to finding the vector of coefficients $\theta$, so that

$$y = \theta^T x$$

subject to minimizing the error function

$$J(\theta) = \frac{1}{2} \sum_i (\theta^T x^i - y^i)^2$$

where the subscript $i$ refers to a specific training sample.

This problem can be solved recursively with the least mean squares (LMS) update rule:

$$\theta_j = \theta_j + a(y^i - h_\theta(x^i))x^i_j$$

for each coefficient component $\theta_j$ (with stochastic or batch gradient-descent) or analytically, obtaining the coefficients directly via the formula

$$\theta = (X^T X)^{-1} X^T y$$

The multiple linear regression term refers to the fact that $x$ is a vector. The method can be extended in the case where the mapping function does not have linear dependencies only, but it can also contain (non-linear) functions of the input. For example, for a one-dimensional input, cubic regression takes the form:

$$y = \theta_0 + \theta_1 x + \theta_2 x^2 + \theta_3 x^3$$

The second component of our method, the artificial neural network (ANN)[29, 32], consists of (a single or multiple layers of) neuron units. The neuron is the building block of the ANN and it is show in Figure 5b. It consists of a linear combiner which sums the weighted inputs and adds a constant bias term and a non-linear unit, which corresponds to the activation function, that takes as input the linear combiner output and applies the nonlinear function $f$ to it. Examples of non-linear functions frequently used are the sigmoid, the hyperbolic tangent and the radial functions.

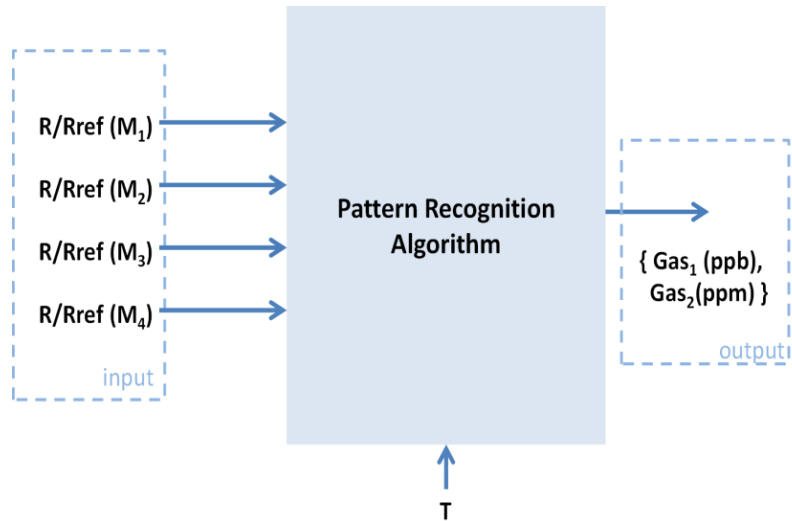

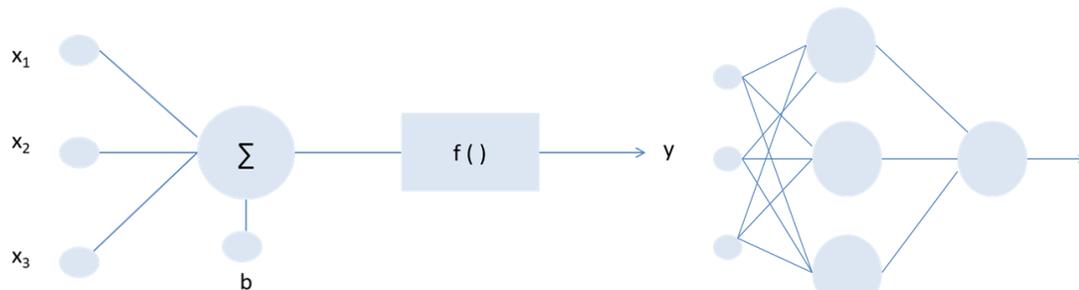

**Fig. 5 (a)** Mathematical setup of the pattern recognition problem **(b)** Components of the hybrid algorithm: Neuron model with linear combiner and nonlinear unit **(c)** Components of the hybrid algorithm: ANN with one hidden layer consisting of three neurons

In Figure 5c, we see an example of an ANN, one that consists of 3 layers: input, output and a single hidden layer, which is the topology we will be using in our analysis.

The algorithm used to train the ANN is back-propagation (and its variants). Assuming that the ANN has $M$ layers and the target output is $d$, and denoting the multiplying weights for layer $l$ by $W_{ij}$, with the subscript $i$ corresponding to the layer and the subscript $j$ denoting the neuron, the equations governing the training algorithm are as follows:

$$y_i = x_i^{(l)} = f\left(\sum_j W_{ij}^{(l-1)} x_j^{(l-1)} + b_i^{(l-1)}\right)$$

$$\Delta W_{ij}^{(l-1)} = \mu x_j \delta_i^{(l-1)}$$
$$\Delta b_i^{l-1} = \mu \delta_i^{l-1}$$

where

$$\delta_i^{(l-1)} = \begin{cases} f'(x_i^{(l-1)})(d_i - y_i), & l = M \\ f'(x_i^{(l-1)}) \sum_k W_{ki} \delta_k^{(l)}, & 1 \leq l < M \end{cases}$$

and $\mu$ denotes the learning rate.

In our scheme, we combine the multiple linear regression with the ANN approach, using the temperature as a guiding element. The main advantage of our multi-pass method is that it maintains its simplicity, without compromising the detection accuracy, to the extent possible. Alternatively, it can be used to identify the resolution of gas mixture concentrations that can be identified with the sensor array, given its operational temperatures. We observed that breaking down the detection algorithm in multiple stages improves significantly the prediction accuracy, compared to the accuracy obtained from training the algorithm by aggregating the data from all temperatures. Finally, the multiple linear regression models proved to be accurate for training purposes, at least in the first algorithmic stages. Artificial neural networks are employed when the regression models perform poorly and their purpose is two-fold: they are used both for training, as well as for defining the resolution limits of the detection algorithm, given the available data.

For each of these steps a (potentially) different set of data (coming from a different temperature) is used to achieve the best classification accuracy. Note that the hybrid neural network is employed only when the classification results are 'out-of-range', which, in this context, means that we need to go beyond multiple linear regression methods to detect the particular gas range, as we have reached the limits of the regression. The process is repeated for all gasses of interest. Alternatively, the algorithm can be used to define the optimal concentration ranges for detection given a certain confidence level. In this manner, the simplicity of the algorithm is preserved, to the extent possible, and the gas mixture components can be modeled independently; which makes the method modular and easy to generalize. Specific implementations of this algorithm are shown in the next sections. A prototype code in R is used for the analysis that follows[33].

**Detection Results for $O_3$**

We consequently used a multiple regression model to predict the gas' concentrations from the four (mean) metal-oxide resistance values. We model each gas separately, using cubic interpolation on the four input variables, for the (first) algorithmic stages. We should also mention that the temperatures used for each pass are the optimal ones for the mixture detection from the measured data. We found that this multi-pass approach is more accurate than the aggregate approach where one model is used to fit the data from all temperatures and concentrations.

We will break down the detection of $Gas_1$ ($O_3$) in two stages, where the temperature range defines the different algorithmic stages (passes). The first pass is based on data from 200°C. In this first stage, we try to determine whether the $O_3$ concentration is above or below a given threshold, set at 400ppb. The second stage consists of two sub-stages, based on the results of the first pass: if the sample is below 400ppb, the data collected at 300°C will be used to fit the lower concentrations, otherwise the data from 200°C will be used to predict the exact concentration for values greater or equal to 400ppb. For $O_3$, we will not augment the algorithm with an ANN, as the results are very accurate with the two-stage regression. It is also noteworthy how different temperatures augment the gas detection of different concentration ranges.

To better illustrate the process, we present the results of the training phase (training error), and, after the methodology overview is completed, we also show the results of the leave-one-out cross-validation process.

With respect to the first stage results, where the data from the lowest temperature (200° C) are used to classify the gas ($O_3$) concentrations in two ranges with a threshold value of 400 ppb, we can achieve almost perfect accuracy, in terms of training error. Once these two ranges have been identified, the data from 200° C are used to detect the exact gas concentration, for the higher concentration ranges. We found that the data from 200° C are not efficient in telling apart the lower concentrations. To fit the lower values, we use the data from 300° C, getting excellent training accuracy (Figure 6a). The adjusted $R^2$ values for these two fits are 0.9792 and 0.9901, respectively.

We can map this regression problem to an equivalent classification one, by truncating the estimated values to the closest (discrete) concentration value of the experiment, i.e. by using as truncation threshold the midpoint of consecutive actual concentration ranges. In this case, the truncation error in the training phase, for the first pass is 1/90; i.e. we misclassify one sample out of the 90 available data-points, when we decide whether the sample in question is above or below 400ppb. The effect of the first-pass decision and the error propagation will be quantified better in the future with more trials.

We repeat this process, reporting the accuracy on a test set, using leave-one-out cross validation, due to the small dataset (90 points), for all measurements. This analysis gives us a more reliable error estimate, as the training error may be too optimistic. Throughout this analysis, we truncate the estimated values to their nearest actual value, using as threshold the midpoint between two consecutive ranges and we take into consideration the error propagation, by omitting from the training set examples that were misclassified in the first pass. We repeat the process leaving out every time a different sample, until all samples are exhausted; therefore, we use 89 samples for training and we leave 1 out, repeating the process 90 times. The collective results of this analysis are shown in Table I.

**Table I:** Cross validation results for $O_3$

| Stage | Range (ppb) | Threshold | Errors |
|---|---|---|---|
| 1 | [0, 800] | 400 ppb | 1 |
| 2_a | [400, 800] | Midpoint of corresponding ranges | 3 |
| 2_b | [0, 400) | Midpoint of corresponding ranges | 2 |
| No of Samples | | 90 | |
| Accuracy | | 93.3% | |

Specifically, using leave-one (trial)-out cross validation, the first pass for $O_3$ when we decide whether the gas concentration is smaller than 400 ppb or not, we only have 1 miss (1/90). For the second pass for the identification of the higher concentrations, we misclassify 3 samples in this subset, and for the lower range we only fail to recognize 2 samples, after repeating this process 90 times, leaving every time a different sample out. Therefore, the detection accuracy is ~93% for $O_3$, or equivalently, the generalization error is close to ~7%.

**Detection Results for CO**

The methodology for the CO gas detection follows the same multi-stage regression algorithm, augmented by a neural network. As in the case of $O_3$, the temperature combinations reported here are the ones that produce the most accurate results for the current dataset. In a similar manner to the $O_3$ detection, in the first stage, we use the data collected at 400° C to decide whether the data sample corresponds to a concentration smaller or greater than 25 ppm. The second stage consists of two sub-stages, in the first of which we will only fit the larger concentrations (greater than 25ppm) and let the smaller ranges be identified with a multi-stage artificial neural network (ANN), as we cannot achieve the desired accuracy with a regression model. The stages and the decision boundaries of the ANN that define the corresponding detection concentration ranges are as follows: using data collected at $200°C$, we will set the threshold at 10ppm and the lower concentrations with respect to this threshold will be further processed to decide on the presence (or absence) of CO, from data at $400°C$.

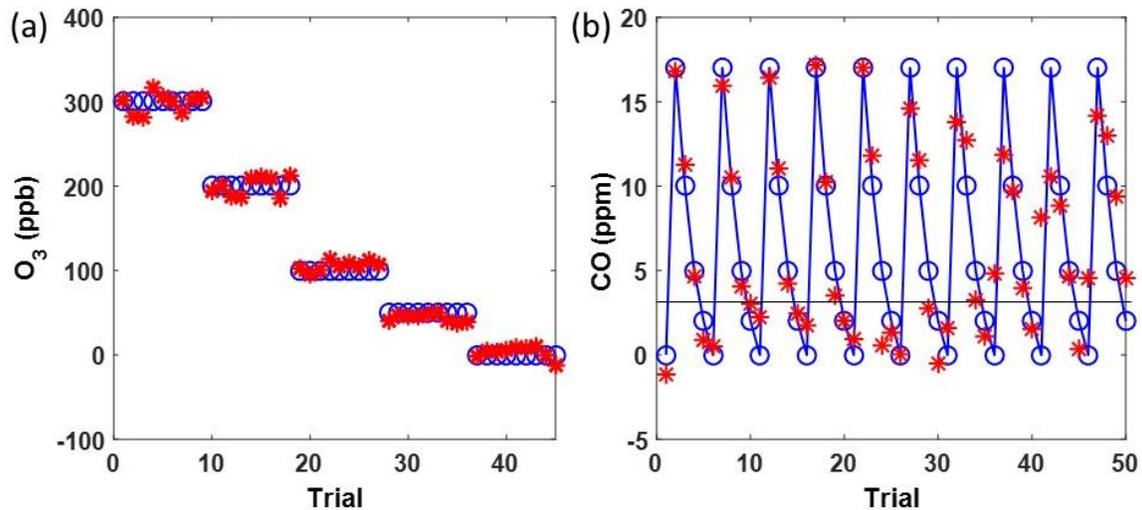

**Fig. 6:** Examples for training values and comparison between estimated (red) and actual (blue) values for **(a)** for $O_3$ concentrations smaller than 400 ppb at 300° C and **(b)** 25ppm for CO concentrations smaller than 25 ppm at 200° C.

For the regression part of the problem, the fit for the higher concentrations in the training set is very accurate with an adjusted $R^2$ value of 0.9597. However, the fit for the lower concentrations is not very accurate with an adjusted $R^2$ value of 0.7771. So, instead of regression we will train an ANN to fit the lower concentration ranges of CO.

After mapping the regression problem to the corresponding classification one, by truncating the predicted values to the closest actual concentrations and repeating the leave-one-out cross validation process, until all the samples have been used as testing points, the collective worst-case results are shown in Table II.

**Table II:** Cross validation results for CO

| Stage | Range (ppm) | Threshold (ppm) | Values (ppm) | Errors |
|---|---|---|---|---|
| 1 | [0, 50] | 25 | [0,25), [25,50] | 7 |
| 2_a | [25, 50] | Midpoint of corresponding ranges | {25,37,43,50} | 3 |
| 2_b_i | [0, 25) | 10 | {0,2,5}, {10,17} | 1 |
| 2_b_ii | [0,10) | [0] | {0},{2,5} | 4 |
| No of Samples | | 90 | | |
| Accuracy | | 83.3% | | |

In particular, the first pass when we decide whether the gas concentration is less than 25 ppm or not, leads to a total of 7 misses out of 90 samples and 3 misses for the second pass for the higher range classification. The number of samples is 90, since all these data points correspond to mixture measurements. To the last part of the second stage, we will train an ANN to augment the detection mechanism for concentration values smaller than 25ppm. Thus, in the CO detection, the output of the first pass for the lower ranges (concentration smaller than 25ppm), will be fed into a multi-stage neural network, as follows.

With respect to the lower range results, throughout this analysis, we will use a three-layer artificial neural network (ANN)[32] with one hidden layer with four neurons. The logistic function was used as the activation function, along with the modified globally convergent back-propagation algorithm and radial basis functions.[32, 34] We try to find the optimal ranges (classes) that would give us the smallest prediction error. Each time we solve a binary problem, in a divisive clustering approach[30], thus solving the problem hierarchically with a multiple-stage neural network structure.

In particular, the classes ($Gas_2$ concentrations in ppm) chosen for the first clustering step are the following: Class 1 consisting of concentrations {0, 2, 5} ppm and Class 2 consisting of concentrations {10, 17} ppm. The neural network(s) using data collected at 200° C managed to distinguish successfully between these two classes, missing only 1 case, which was a lower concentration misinterpreted as a higher one. Dividing further Class 1 into two subclasses using data collected at 400° C, we can tell concentrations {0} ppm from {2, 5} ppm apart, missing 4 cases on this subset, which were all false alarms. All the results reported are on the test-set with leave-one-out cross validation, repeated until all data points are exhausted (Fig. 6(b)).

These are the optimal subdivisions for this dataset that give satisfactory accuracy based on the optimal data combination, in terms of temperature ranges and concentration bins. Thus, we conclude that we can successfully discriminate among most concentrations for $Gas_2$ with (hierarchical) neural net modeling, but we may need to concatenate adjacent ranges in certain cases. It is expected that the accuracy will increase with increasing representative datasets.

With this algorithm, the final leave-one-out cross results for the binary mixture of ($O_3$ -CO) are: ~93.3 % accuracy was achieved for $Gas_1$ ($O_3$) and 83.3% accuracy was achieved for $Gas_2$ (CO) and ~88% average accuracy for the binary mixture.

Finally, in Table III, we emphasize the dependency of the detection ranges on the temperature for the two gases of our mixture. From this table it becomes evident that different operating temperatures of the array lead to different detection ranges. It should also be noted, that increasing the number of operating frequencies may improve the concentration resolution.

**Table III:** Dependency of detection values on temperature

| $O_3$: Range (ppb) | Temperature (°C) | CO: Range/Classes (ppm) | Temperature (°C) |
|---|---|---|---|
| [0,400), [400,800] | 200 | [0,25), [25,50] | 400 |
| [0,400) | 300 | {25,37,43,50} | 200 |
| [400,800] | 200 | {0,2,5}, {10,17} | 200 |
|  |  | {0},{2,5} | 400 |

As future work, as we collect more data, it would be worth investigating the effect of (temperature-dependent) principal component analysis (PCA)[29,31] in identifying clusters in an unsupervised fashion, as a pre-stage in our analysis. These clusters can subsequently be fed into the remaining flow of our algorithm.

Furthermore, the effect of the error propagation should be thoroughly studied and will be quantified in detail in the future.

It would also be interesting to see the resolution limits of the sensor array, as determined by the minimum accuracy threshold and define the optimal detection ranges. In cases where greater resolution is necessary, the existing system can be augmented by the introduction of more temperatures to facilitate the detection process.

Finally, with the introduction of more gases, it would be worth investigating whether a pre-classification stage should be added, that would correspond to the mere identification of the mixture components, before entering the concentration detection phase.

# CONCLUSIONS

In summary, we have microfabricated a monolithic sensor array comprised of 4 metal oxide semiconductor materials. The architecture of the arrays is optimized to minimize heat dissipation, resulting in an average power consumption per pixel of only 27μW at 300° C operating temperature. We have also resolved the individual components within a homogeneous mixture of gases. The array principle was applied to $O_3$-CO mixture wherein a library of concentration combinations at various operating temperatures was used to train a hybrid novel machine learning algorithm which was then used to predict component gas concentrations with high accuracy. Although shown for a binary mixture, the methodology can be extended to multiple gases and VOCs. Datasets that take into account the sensor pixel aging and different environmental conditions such as humidity which are known to affect the sensitivity of MOS sensors can also be used to train the hybrid regression/ANN algorithm to enable predictions under such conditions. Such a system can provide a self-calibrated solution that would enable sensitive and selective gas sensing that meets the requirements of wearable/portable solutions.

# ACKNOWLEDGEMENTS

The authors would like to thank their managers, M. Huang, X. Su and K. Foust for their feedback and support on this work.

# COMPETING INTERESTS

The authors declare no conflict of interest.

**Supporting Information for**

**Metal-Oxide Sensor Array for Selective Gas Detection in Mixtures**

*Noureddine Tayebi[†]\*, Varvara Kollia[†]\* and Pradyumna S. Singh\**

\*Intel Labs, Intel Corporation, 2200 Mission College Boulevard, Santa Clara, CA 95054, USA

(† Work reported herein was done when these authors were at Intel.)

# S1. Sensor Array Fabrication

The fabrication process of the metal-oxide sensor array is as follows. First, silicon wafers are diffusion clean (Fig. S1-Step 1), which is followed by the deposition of a 500 nm thick low stress silicon nitride is deposited (Fig. S1-Step 2). The silicon nitride film was deposited using low-pressure chemical vapor deposition at 250 mTorr engineered to result in very low stress membranes. This is critical to lower susceptibility to damage by repeated temperature cycling. A 100 nm thick platinum layer is then deposited and patterned by lift-off to form serpentine features which serves as heating and resistive (for temperature sensing) elements (Fig. S1-Step 3). This layer is then passivated by a 200 nm silicon nitride film (Fig. S1-Step 4). This is followed by the deposition and patterning through lift-off of another 100 nm thick platinum which in this case form the sensing interdigitated electrodes (Fig. S1-Step 5.) The various metal oxides are then individually deposited and patterned on each pixel location using reactive sputtering (Fig. S1-Step 6). Finally, an undercut is created under the active area of each pixel. This is achieved by first opening windows within the silicon nitride lithographically and etching silicon nitride in a CHF3 chemistry. This is then followed by etching the exposed silicon substrate areas (Fig. S1-Step 7).

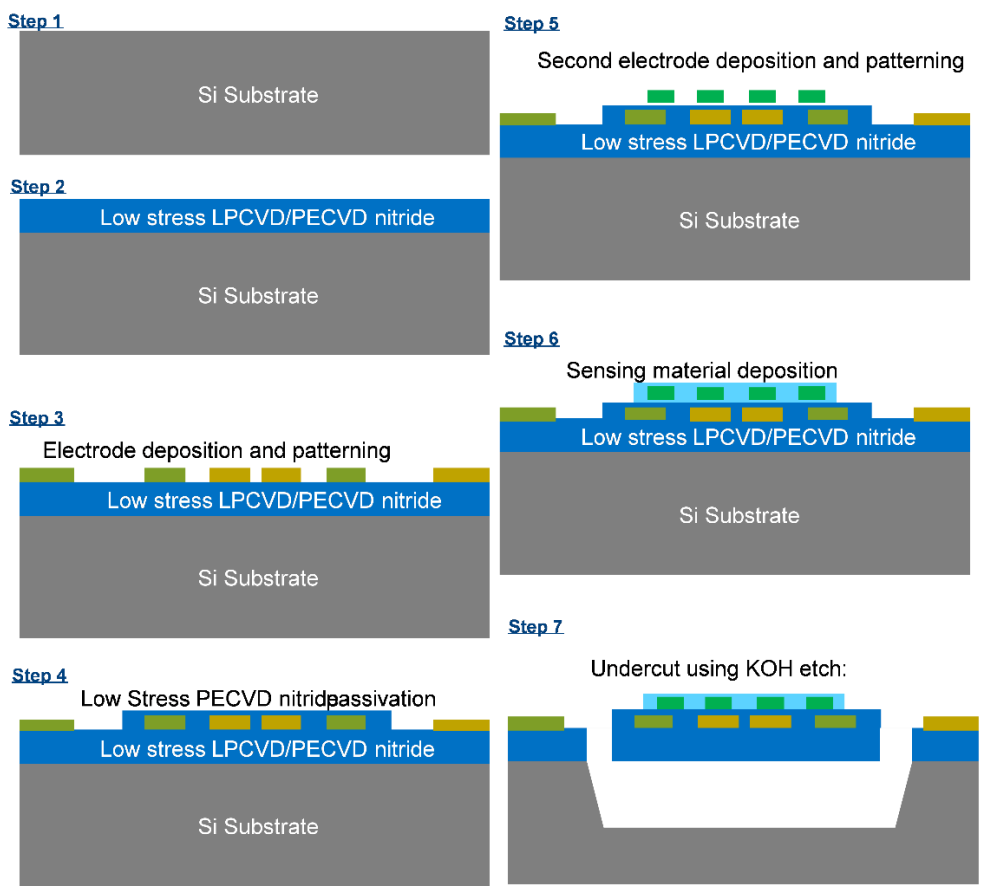

**Fig. S1.** Microfabrication process steps of the metal oxide sensor array.

# S2. Gas Delivery and Testing set-up

We custom built a gas testing chamber as outlined in the figure below (Fig S2). Carbon Monoxide (CO) gas was delivered to the device using CO cylinders (GASCO, Calibration gas, 50 ppm in Air). Ozone was generated using an Ozone generator (Analytical Instruments, PA) connected to a 20.9% O2/N2 cylinder (GASCO, Calibration gas).

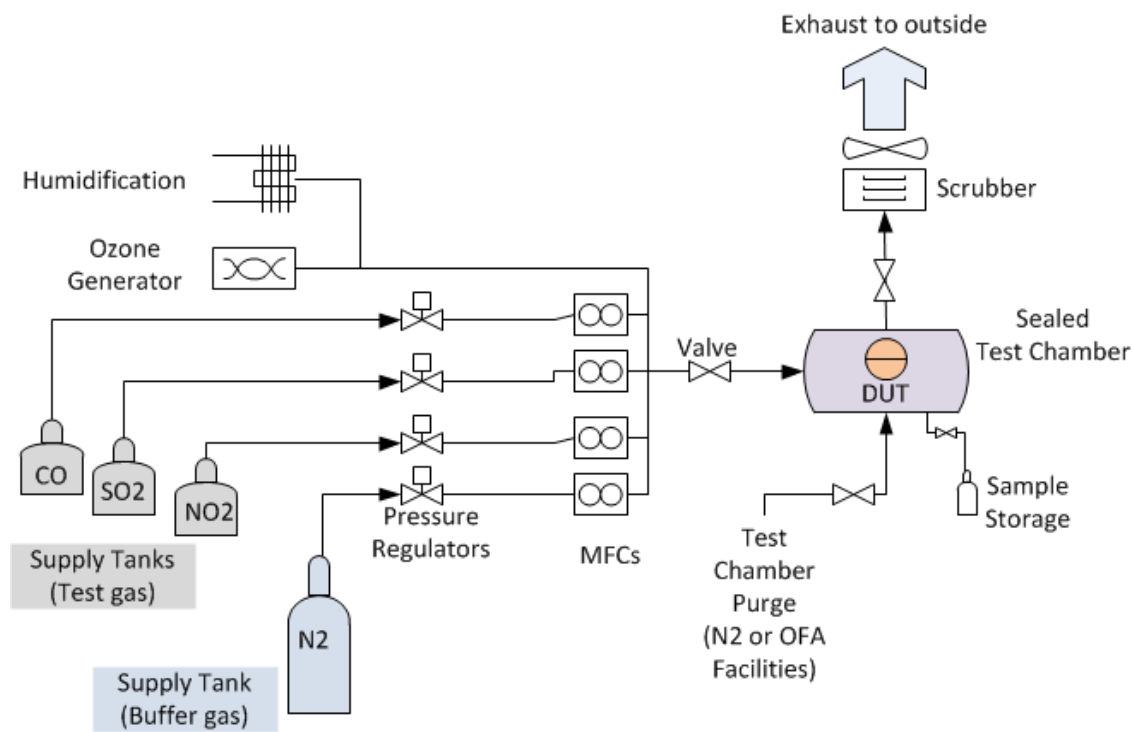

**Fig. S2.** Schematic of the custom-built gas sensor testing set-up.

The ozone was delivered at a fixed flow rate of 500 SCCM. Oil-free Air was used for dilution purposes. The desired concentrations were set by controlling the flow rates of the individual gases using mass flow controllers (Brooks Instruments, GF series and 4800 series).The devices were addressed using a custom designed printed circuit board (PCB) connected to a Keithley 4200 Semiconductor analyzer.